\documentclass[runningheads]{sig-alternate-10pt}
\usepackage{makeidx}  % allows for indexgeneration
\usepackage{epsf}\epsfverbosetrue
\usepackage{psfrag}
\usepackage{indentfirst}
\usepackage{subfigure}
\usepackage{wrapfig}
\usepackage{amstext}
\usepackage{amsmath}
\usepackage{amssymb}
\usepackage{amsfonts}
\usepackage{times}
\usepackage{url}
\usepackage{graphicx}

\def\bt{BitTorrent}
\def\p2p{peer-to-peer}
\def\P2p{Peer-to-peer}

\def\ih{infohash}

\def\ut{$\mu$Torrent}

\begin{document}

\title{Compromising Tor Anonymity \\ Exploiting P2P Information Leakage}

\author{
Pere Manils, Abdelberi Chaabane, Stevens Le Blond,  \\
Mohamed Ali Kaafar, Claude Castelluccia, Arnaud Legout, Walid Dabbous \\
}

 \maketitle
\sloppy

\begin{abstract}

Privacy of users in P2P networks goes far beyond their
current usage and is a fundamental requirement to the adoption of
P2P protocols for legal usage. In a climate of cold war
between these users and anti-piracy groups, more and more users are
moving to anonymizing networks in an attempt to hide their
identity. However, when not designed to protect users information, a
P2P protocol would leak information that may compromise the identity
of its users. In this paper, we first present three attacks targeting \bt{} users on top
of Tor that reveal their real IP addresses. In a second step, we analyze the
Tor usage by \bt{} users and compare it to its usage outside of Tor. Finally, we depict
the risks induced by this de-anonymization and show that users' privacy violation
goes beyond \bt{} traffic and contaminates other protocols such as HTTP.

\end{abstract}

% Introduction
\section{Introduction}

%%%%%%%%%%%%%%%%%%%%%%%

%%%%%%%%%%%%%%%%%%%%%%%

The Tor network was designed to provide freedom of speech by
guaranteeing anonymous communications. Whereas the cryptographic
foundations of Tor, based on onion-routing \cite{onionrouting0,onionrouting1,onionrouting2,onionrouting3}, are known to be 
robust, identity leaks at the application level can be exploited by adversaries to
reveal Tor users identity. Indeed, Tor does not cipher data streams end-to-end, but from the
source to a Tor exit node. Then, streams
from the Tor exit node to the destinations are in plain
text (if the application layer does not encrypt the data).
Therefore, it is possible to analyze the data stream seeking for
identity leaks at the application level. Tor does not consider
protocol normalization, that is, the removal of any identity leak at the
application level, as one of its design goals. Whereas this assumption
is fair, Tor focuses on anonymizing the network layer, it makes the
task of users that want to anonymize their communications much harder. As an illustration, the Web communications on Tor are the subject of
many documented attacks. For instance, attacks can leverage from misbehaving browsers to third party plugins or web components  (JavaScript,
Flash, CCS, cookies, etc.) present in the victim's browser to reveal
browser's history, location information, and other sensitive data \cite{tor_http_attack1,tor_http_attack2,tor_http_attack3,tor_http_attack4}.

In order to prevent or at least reduce these attacks, the Tor project recommends the use of web proxy solutions like Privoxy or Polipo \cite{privoxy,polipo,recomendation_privoxy}. The Tor project is even maintaining a Firefox plugin (Torbutton \cite{torbutton}) that, by disabling potentially vulnerable browser components, aims to counter-measure most of the well-known techniques an adversary can exploit to compromise identity information of Tor users. Thus a big effort has been invested and is heading on improvement and protection of the HTTP protocol on top of Tor, but surprisingly, such an effort is limited to this protocol.

P2P applications and more specifically \bt{}, an application that is being daily used by millions of users~\cite{stevens_spying}, have been so far neglected and excluded from anonymizing studies. The crux of the problem is that \bt{} easily allows any adversary to retrieve users' IP addresses from the tracker for torrents they are participating to. Indeed, by design \bt{} exposes the IP address of peers connected to torrents. This implies important anonymity and privacy issues, as it is possible to know who is downloading what. To go around this issue, many \bt{} users that care about their anonymity use Tor, although the Tor project explicitly not recommend the use of \bt{} on top of the Tor network, because of the major risk of overloading the network.

\bt{} is a complex protocol with many potential identity leaks, as user privacy is not among its design goals. However, this serious issue is overlooked by \bt{} users who believe that they can hide their identify when using Tor.

Today's reality is that \bt{} is one of the most used protocols on top of Tor (with HTTP/HTTPS) in terms of traffic size and number of connections as reported by \cite{Shining_Dark} and observed during our own experiments. Surprisingly, no studies have been conducted on the way \bt{} may leak the identity of users when the application is running over an anonymizing network. Although, it might be argued that \bt{} is mostly used for piracy (distribution of illegal content), we believe that privacy
issue is a major impediment for the commercial and legal use of
\bt{}. Moreover, identity leaks at the level of a stream might
also contaminate other streams, thus compromising non-\bt{} traffic.

Our study attempts to answer the following three questions:

\begin {itemize}
\item How is Tor being used by \bt{} clients?
\item Does the anonymity and privacy's vulnerability of \bt{} makes Tor less anonymous, leaking information about other Tor usage?
\item To what extent can we use de-anonymization to track users and break their privacy through Tor?
\end{itemize}

The answers to these questions have implications in numerous Tor security applications. In essence, we show in this paper that there is a gap between users' willingness to use \bt{} anonymously and their expectation to hide their Internet activities through Tor.

As a first contribution (Section~\ref{sec:attacks}), we show using three techniques, how a malicious exit node may de-anoymize \bt{} traffic. Two of our proposed techniques are completely passive, relying on information leakage of the application itself (in our particular case, it is the \bt{} client leaking information). The third technique is active and exploits the lack of authentication in the \bt{} protocol.

As a second contribution (Section~\ref{sec:attack_stats}), we demonstrate using a so-called \emph{domino effect} that the identity leak contaminates all streams from the same Tor circuit,  and, even from other Tor circuits. In particular, we show that \bt{} users' privacy may be infringed, and more importantly, that these privacy issues may go far beyond P2P traffic.

Exploiting our attacks, we provide as a third contribution (Section~\ref{sec:BtUsage}) the first in-depth study of \bt{}'s usages on top of Tor. In particular, we quantify how \bt{} users interact with Tor, and detail their behaviors compared to regular \bt{} users.

Finally, we show how our attacks can be used to perform profiling of \bt{} users (Section~\ref{nonAnonymat_privacy}). Focusing on the HTTP protocol (being the most used and most protected by Tor) we show that a significant quantity of information is leaking, proving that we can quickly move from a P2P anonymity weakness on top of Tor to privacy issues.

As a conclusion,with hope that our work will contribute to the ongoing debate on the balance between anonymity and privacy preserving and performance-efficient applications (e.g. \cite{letter}), we show that the fixes of the anonymity issues we identified may involve support of different cryptographic operations between \bt{} entities, particularly when used on top of Tor.

% Background
\section{Background}
\label{sec:background}

In the following, we provide a brief overview of the Tor anonymizing
network. We also introduce different aspects of the \bt{} protocol, being largely exploited in our attacks.

\subsection{Tor Overview}
\label{sec:tor-overview}

Tor is a circuit-based low-latency anonymous communication service
\cite{Dingledine04tor}. Its main design goals, as stated in the
original paper, are to prevent attackers from linking communication
partners, or from linking multiple communications to or from a single
user.

Tor relies on a distributed overlay network and on onion routing
\cite{onionrouting0} to anonymize TCP-based
applications like web browsing, secure shell, or P2P
communications.

When a client wants to communicate to a server via Tor, he selects $n$
nodes of the Tor system (where $n$ is typically 3) and builds a
\textit{circuit} using those nodes. Messages are then encrypted $n$
times using the following \textit{onion encryption} scheme: messages
are first encrypted with the key shared with the last node (called
\textit{exit node}) on the circuit, and subsequently with the shared
keys of the intermediate nodes from $node_{n-1}$ to $node_1$. As a
result of this onion routing, each intermediate node only knows its
predecessor and successor, but no other nodes of the circuit. In
addition, the onion encryption ensures that only the last node is able
to recover the original message.

Onion routing originally built a separate circuit for each TCP stream.
However, this required multiple public key operation for every
request, and also presented a threat to anonymity from building so
many circuits \cite{Dingledine04tor}. Tor, instead, multiplexes
multiple TCP streams from the same source on a single circuit to
improve efficiency and anonymity.  In order to avoid delays, circuits
are created preemptively in the background.  Also to limit
linkability, new TCP streams are not multiplexed in circuits
containing already older-than-10-minutes streams.

A Tor client typically uses multiple simultaneous circuits.
As a result, all the streams of a user are multiplexed over these
circuits.  Thus, connections to the tracker and connections to the peers
 can be assigned to different circuits.

\subsection{\bt{} Information Leakage}
\label{sec:bt-info-leak}

\begin{figure}[!t]
  \centering
  \includegraphics[width=1\columnwidth]{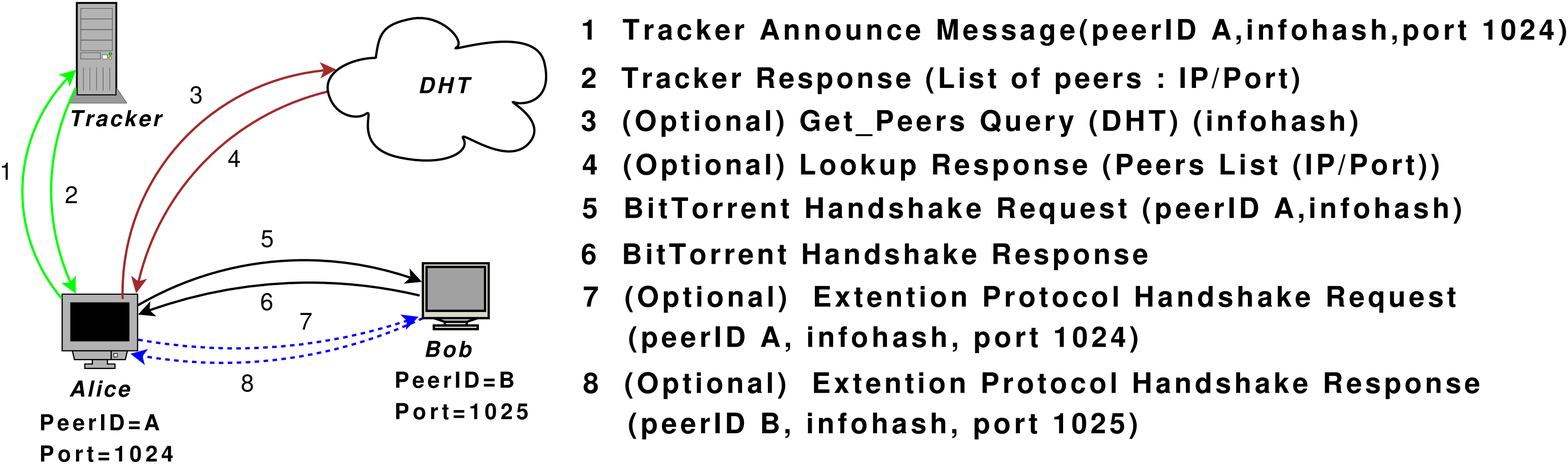}
  \caption{\bt{} Protocol Diagram}
  \label{fig:bt_diag}
\end{figure}

A torrent is a set of peers sharing the same content. In this section,
we briefly describe the information that can leak from a peer Alice
when she joins a torrent as in Figure~\ref{fig:bt_diag}.

To join a torrent, Alice sends an \textit{announce} message to the
tracker that maintains the list of all peers in that torrent (step 1
in Figure~\ref{fig:bt_diag}). The announce message is an HTTP GET
message that contains three important identifiers that we used in our
attacks: i) The \ih{} that is a 160 bits unique identifier of a
torrent. ii) The TCP port number selected randomly at the installation
of the client on which the peer is listening on. iii) The peer ID of
the client, that is the concatenation of an identification of the
client version and a random string. This peer ID can be generated at
client installation, each time the client is restarted, or each time
Alice joins a torrent.  iv) Optionally, the IP address of the
interface from which Alice sent the message.
Once the tracker receives the announce message for a specific torrent
identified by the \ih{}, it selects a random subset of peers in that
torrent and returns the endpoints (the IP and port of a peer) of those
peers (step 2).

Alice can also use a DHT \cite{MAYM02_IPTPS, bep5} that runs on top
of UDP, to find peers in a torrent. In order to retrieve the list of
peers, Alice performs a \textit{find\_node} query
containing the \ih{}. The result of this query is the ID of the DHT
node that maintains the tracker for the queried \ih{}. Then, Alice
performs a \textit{get\_peers} query to the DHT node in order to
retrieve the endpoints for a random subset of the peers already in the
torrent (steps 3 \& 4). As with a tracker, Alice can retrieve all the
endpoints\footnote{By performing numerous queries to the DHT.} of a torrent with the DHT.
Then, Alice establishes a TCP connection and sends a handshake message
to each peer (steps 5 \& 6). This handshake message contains the \ih{}
of the torrent, and the peer ID. The port number the peer is listening
on is present in the handshake when the extended messages
option~\cite{bep10} is enabled in the \bt{} client (steps 7 \&
8). This is the case by default with \ut{}, the most popular
BitTorrent client.  The extended handshake, {\it sometimes},
contains the IP address of Alice. We will come back to this issue in
Section~\ref{sec:attack1}.

Finally, popular BitTorrent clients, e.g., \ut{} and Vuze, allow to
configure SOCKS proxies and give the option to use the proxy for
connections to the tracker, to the peers, or both. Therefore, a
BitTorrent client can use Tor, configuring the Tor interface as a
SOCKS proxy, for communication to the tracker or the peers
independently. Alice can then decide to connect to the tracker via
Tor, but to have a direct connection to peers in order not to have
performance penalty.

% Attacks
\section{De-anonymizing BitTorrent Users}
\label{sec:attacks}

In the following, we describe the experimental methodology and techniques used to de-anonymize \bt{} users on top of Tor, and we present the results of their evaluation in the wild.

% Methodology and experimentation
\subsection{Methodology}
\label{sec:methodology}

To de-anonymize the IP address of \bt{} users in the wild, we instrumented and monitored $6$ Tor exit nodes for a period of $23$
days. From January $15$ to February $7$th, we monitored the Tor
traffic on controlled exit nodes that were distributed around the globe: two
in Asia, two in Europe, and two in the U.S. As anyone can volunteer to
host a Tor exit node, performing the attacks described in this paper is within any adversary's grasp. 

In order to comply with the legal and ethical aspects of privacy, we
performed our analysis on the fly. In addition, special cautionary measures were taken in order to present only
aggregated statistics as suggested by Loesing et al. in~\cite{loesing:financialcrypto2010}.

% Attack 1
\subsection{Simple Inspection of BitTorrent control messages}
\label{sec:attack1}

In this section, we show that an attacker can de-anonymize
the IP of a \bt{} user simply by looking at the IP field
contained in the \bt{} control messages introduced in
Section~\ref{sec:bt-info-leak}, i.e., announce and extended
handshake.

\textbf{Tracker Announces.} As we have mentioned in Section~\ref{sec:bt-info-leak}, the announce
message is sent to the tracker to request a list of peers in a
torrent. Depending on the client, that message may contain the IP address of
the user.

We captured $200$k announce messages on our exit nodes. Among the $35\%$ of
those messages that contained a non-empty IP parameter, $4\%$ were invalid IP addresses, $38\%$ contained a private IP address, and the
remaining $58\%$ contained a public IP address. We ended up with $3,698$
unique public IP addresses.

Surprisingly, most of the public IP addresses we found were IPv6. We
also observed that the same versions of \bt{} clients were alternating
between two behaviors, embedding in some cases public IP addresses and in others private ones.
The top 3
\bt{} clients that were constantly embedding public IP addresses (normalized by
their presence in our traces) were \ut{}, BitSpirit, and libTorrent.

\textbf{Extension Protocol Handshakes.} After a regular \bt{} handshake, a client supporting the Extension
Protocol may send an additional handshake as described in
Section~\ref{sec:bt-info-leak}. That extended handshake message may
also contain the user's IP address.

We captured $45$k extended handshakes on our exits nodes. In $84\%$ of
the handshakes an IPv4 address was present. Of those messages containing an
IP address, $33\%$ contained a public IP address that was not the IP of a Tor exit node.
In total, we collected $1,131$ unique public IP addresses.

In $67\%$ of the handshakes containing an IPv4 address, the IP belonged to an
exit node. Although we have not tested the behavior of those clients,
we suspect that they use a service to determine their IP address as seen from
the Internet. As they will contact that service through Tor, the
service will report the IP address of an exit node.

\textbf{Conclusions.} The inspection of BitTorrent control messages is the simplest of the
three attacks that we identify in this paper. To conduct this attack,
an attacker only needs to monitor the announce and extended handshake
messages on a Tor exit node. However, we have not checked the
authenticity of the public IP address contained in those messages, therefore
we do not include them in our statistics.

% Attack 3
\subsection{Hijacking Trackers' Responses}
\label{sec:attack3}

Unlike the previous attack, hijacking the tracker responses guarantees
that the de-anonymized IP belongs to the BitTorrent user. Assume Bob
is the attacker in Figure~\ref{fig:bt_diag}. Hijacking consists in
rewriting the list of peers returned by the tracker to Alice so that the
first endpoint in the list belongs to Bob. If Alice uses Tor only to
connect to the tracker, but not to connect to peers, then Bob will
see Alice's public IP address. As the IPs of Tor exit nodes are
public, Bob can easily determine whether he has compromised Alice's
public IP. Hijacking is possible because the communication between
peers and trackers is neither encrypted nor authenticated. This is a
typical man-in-the-middle attack.

Another advantage of hijacking over a mere inspection of the extended
handshakes is that it works even when Alice encrypts her communication
with other peers. Indeed, clients such as \ut{} support encrypted
communication among peers so that a third party, e.g., ISP, cannot
identify that the communication belongs to the \bt{} protocol. In that case, an
eavesdropper will not be able to read the IP field of the extended
handshake but Bob will see Alice's public IP address because she will
establish a TCP connection to Bob. Also note that Bob can answer to
Alice's handshake and let Alice send a piece to him to make sure she
is distributing the content.

Hijacking the announce responses on a single exit node for $23$ days,
we were able to collect $3,054$ unique IP addresses, out of which $814$ ($27\%$) belonged to
exit nodes and $2,240$ ($73\%$) were public. We remind that the
hijacking attack works when Alice uses Tor only for tracker communication,

\textbf{Conclusions.} Hijacking the tracker responses allows an attacker to de-anonymize a
user who only connects to the tracker using Tor. In addition to the
code to instrument and monitor the exit node, this attack requires
approximately $200$ lines of code to rewrite the list of endpoints,
which makes it relatively easy to launch. As we will see in
Section~\ref{sec:BtUsage}, more than $70\%$ of BitTorrent users use
Tor only to connect to the tracker, making hijacking quite efficient
to de-anonymize users.

% Attack 2

\subsection{Exploiting the DHT}
\label{sec:attack2}

\begin{figure}[!t]
  \centering
  \includegraphics[height=2in]{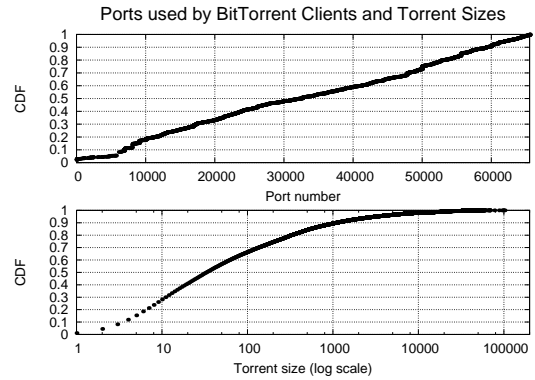}
  \caption{Distribution of the listening ports used by the BitTorrent
    clients that we have de-anonymized on $6$ monitored exit nodes
    during $23$ days (top). Distribution of the torrent sizes in which peers
    using those exit nodes were participating in during the same
    period (bottom). \it{Because the port numbers are uniformly
      distributed and most torrents are small, the port number is a
      good peer identifier.}}
  \label{fig:distr_listening_ports}
\end{figure}

The exploitation of the DHT allows to de-anonymize a user, even if she
uses Tor to connect to other peers. Tor does not support UDP
communications that are used by the DHT. As a \bt{} client will
fail to connect to the DHT using its Tor interface, it connects to the
DHT using the public network interface and publishes its public IP and
listening port into the DHT. Therefore, even though Alice connects to
Bob through Tor, Bob can lookup Alice's public IP address in the
DHT. We have validated this behavior with \ut{}, the most popular
\bt{} client \cite{Zhang09tpds_sub}.

Assume Bob wants to de-anonymize Alice's IP address in the
Figure~\ref{fig:bt_diag} and that Alice sends an announce or extended
handshake message through an exit node that Bob controls. Bob knows
Alice's listening port number and the infohash she's
downloading. Bob can then perform a {\it find\_node} request to find
the tracker of that infohash and iteratively send {\it get\_peers}
messages to it to collect all the endpoints.
If one of the endpoint has the same port as Alice's listening port,
then Bob has most likely de-anonymized Alice's public IP.

We make the assumption that Alice's listening port number is a good
identifier within a torrent. That implies that listening port numbers
are uniformly distributed on $[0;65535]$. As most clients select the
    listening port randomly at the installation of the client, they
    should be uniformly distributed. We confirm that assumption in
    Figure~\ref{fig:distr_listening_ports} (top). However, we exclude
    ports $80$, $443$, $6881$, $16884$, $35691$, and $51413$ that are
    more popular than others. This choice is conservative because we
    accept to de-anonymize less users to reduce the number of false
    positives.

For Alice's listening port number to be a good identifier within a
torrent, that torrent should also be small in terms of size. We also
confirm that assumption in Figure~\ref{fig:distr_listening_ports}
(bottom) where $90\%$ of the torrents have less than $1,000$ peers. By
exploiting the DHT, we de-anonymized $6,151$ unique public IPs.

\textbf{Conclusions.} Exploiting the DHT overcomes the weaknesses of the previous attacks. The de-anonymized IP have a very high probability to belong to
Alice, and it even compromises the IP of Alice if she uses Tor to
connect to other peers. However, as this attack implies to collect all
the endpoints for a given torrent, an attacker should develop a
dedicated crawler or heavily modify an existing client with DHT
support.

In the other hand, not all clients support the DHT. However, the most popular BitTorrent
client, \ut{}, supports it by default. In addition, the current
trend for large \bt{} trackers to promote magnet links
\cite{bep09} instead of torrent files, i.e., DHT instead of trackers,
pushes more and more clients to support the DHT.

% Global statistics on the attacks (dominos)
\subsection{The Domino Effect}
\label{sec:attack_stats}

\begin{figure}[!t]
  \centering
  \includegraphics[height=2in]{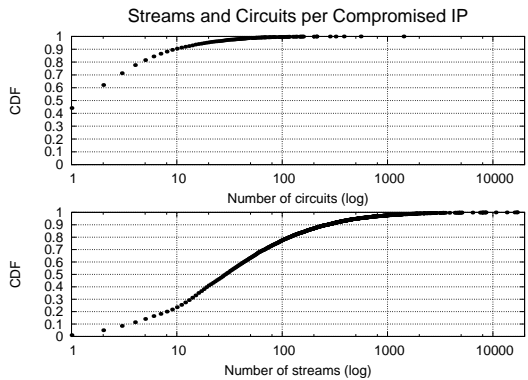}
  \caption{Distribution of de-anonymized circuits (top) and streams
    (bottom) per de-anonymized IP. {\it Once we de-anonymize an IP, we
      are able to de-anonymize multiple circuits and streams using the
      domino effect.}}
  \label{fig:cdf_compromised_streams}
\end{figure}

In the previous sections we have shown different attacks that allow to de-anonymize a \bt{} stream on top of Tor. Furthermore, as
described in Section~\ref{sec:tor-overview}, Tor multiplexes
different TCP streams of the same source over a
single circuit. Consequently, the source of the \bt{} stream
revealed by one of the attacks will also be the source for all the
other streams that are multiplexed in the same circuit as the one
used by the \bt{} stream. We call this issue the
\textit{intra-circuit domino effect}, as identifying the source of a
single stream in a circuit reveals the source for all other streams
multiplexed on this circuit.

 Moreover, a \bt{} stream usually contains the peer ID identifying
 the \bt{} user.  Once a \bt{} stream is de-anonymized, we associate
 the peer ID to the compromised IP address.  Then by simply comparing
 the peer ID of other identified \bt{} streams (belonging to other circuits), we can link the origin of new circuits with
 previously compromised IP addresses, increasing the set of circuits (and thus
 streams) linked to the same user. 
 
 However, the latter approach is not sufficient when the \bt{} traffic is encrypted. The following complementary approach would allow to reveal the IP address of
 the initial source of other encrypted circuits, as long as this
 source has been identified using one of our techniques
 while accessing the tracker (recall that the traffic to the tracker
 is not encrypted). Indeed, first we keep the list of couples
 (IP,port) returned by the tracker to the compromised peer. If, after
 a short period of time (say a few minutes) for a new circuit we
 identify a stream whose destination is one of those (IP,port), we
 deem then the source of this stream is the one previously identified
 by one of our techniques. We call the way we link compromised IP addresses to streams belonging to different circuits, the \textit{inter-circuit domino effect}.

 Whereas this inter-circuit effect allows to identify the source of a
 circuit for which our techniques cannot be performed, it might lead
 to false positives as different peers may share the same list
 returned by the tracker. The choice of a very short period of time
 and the low probability that two different peers choose the same
 exit node to contact the same peer allows then to limit the number
 of false positives.

Figure~\ref{fig:cdf_compromised_streams} shows the CDF of
de-anonymized circuits and streams per IP address. We de-anonymized a few
circuits for most IPs, i.e., less than $10$ circuits for $90\%$ of the
IPs, but a significant number of streams. For $75\%$ of the cases, more than $10$ streams were de-anonymized.
Finally, we observed that the intra-circuit domino effect de-anonymized approximately $80\%$ of the streams, while the remaining  
 $20\%$ were de-anonymized by the inter-circuit domino effect.

% BT usage on tor
\section{\bt{} Usage on Top of Tor}
\label{sec:BtUsage}

Characterizing a real deployed anonymizing network is important. In
particular, McCoy et al. \cite{Shining_Dark} characterized the usage of Tor two years ago and tried to analyze how Tor
is used and mis-used. While these results revealed useful statistics about Tor
usage in general, they did not focus on P2P protocols. Additionally, they did not de-anonymize users and hence, they could not
link particular usages to locations.

In \cite{Shining_Dark}, authors have already shown the importance of
the \bt{} protocol on top of Tor in terms of traffic size and number
of connections. \bt{} ranked the second position among all the
identified protocols, representing $40$\% of all the observed traffic
at an exit node.

Our techniques have revealed many IP addresses of \bt{} users on top of Tor,
providing us with a set of \textit{unique} \bt{} users. More
importantly, once de-anonymized, the IPs may be linked to single
users based on the connections they established on top of Tor. We
first exploited these IP addresses to draw a better view of the individual
usage of \bt{} on top of Tor for each single user. Then, we extended
this information to the whole set of IPs. This allows us to compare
our results with the regular usage of \bt{} outside of Tor, using the traces
collected in~\cite{stevens_spying}.

\subsection{Typical Usage of Tor by \bt{} Users}

\begin{figure}[!t]
  \centering
  \includegraphics[height=2in]{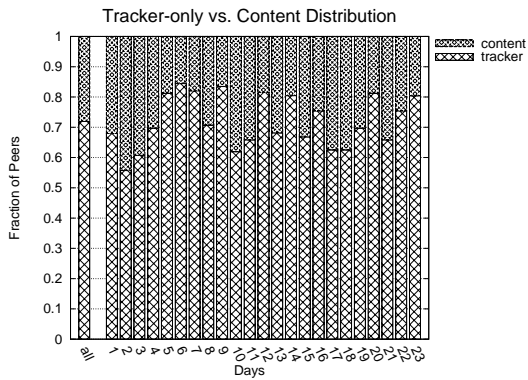}
  \caption{For each day, this histogram shows the proportion of peers
    who use Tor for content distribution (content) or only to connect
    to the tracker (tracker). {\it{all}} is the average over all
    days. \it{$72\%$ of peers use Tor only to connect to the
      tracker.}}
  \label{fig:users-behavior}
\end{figure}

Tor can be used by a \bt{} user to (1) hide from the tracker, (2) hide
from other peers, i.e., content distribution, or (3) hide from both
the tracker and other peers. In this
section, we characterize the usage of \bt{} users on top of Tor.

Usage (1) is the one advocated by the Tor Project in its conditions of
utilization. As \bt{} content distribution overloads the Tor network, the Tor
Project considers usages (2) and (3) as undesirable.

However, it is tempting for users willing to trade performances for
anonymity to use Tor for content distribution thus violating Tor's
conditions of utilization. Quantifying the fraction of users
distributing content over Tor is important for two reasons. First, it
tells the reason why \bt{} users are on top of Tor. Second, it says
how many \bt{} user are responsible of overloading the Tor network.

To quantify the fraction of \bt{} users using Tor for content
distribution, we rely on the hijacking attack. That attack forces a
peer to unwillingly connect to an attacker. As mentioned in
Section~\ref{sec:attack3}, an attacker can easily determine the usage
of a hijacked peer. In particular, a peer with usage (1) will connect
to the attacker from a public IP  whereas a peer with usage (2) or (3)
will connect to the attacker from the IP address of an exit node. We remind
that the IPs of the exit nodes are public so it is easy to determine
whether a peer only hides from the tracker or also from the peers. We
rely on the peer IDs to count the number of unique peers that connect
to us every day.

One limitation of our methodology is that we cannot distinguish
between usage (2) and (3). However, we argue that usage (2) should be
marginal as it implies that a user goes into the trouble of
distributing content over Tor whereas her public IP address is published into
the tracker.

We show the distribution of the peers with usage (1) (tracker-only) and usage (3) (content) in Figure~\ref{fig:users-behavior}. Most
\bt{} users (72\%) only hide from the tracker and do not distribute content
over Tor therefore they respect Tor's conditions of utilization. This
trend is relatively constant in time for a period of $23$ days. As the
peers who only hide from the tracker just send a few announce messages
on Tor every $10$ minutes, this result implies that only few peers are
responsible of most of the \bt{} traffic on Tor.

\subsection{Returning Users}

\begin{figure}[!t]
  \centering
  \includegraphics[height=2in]{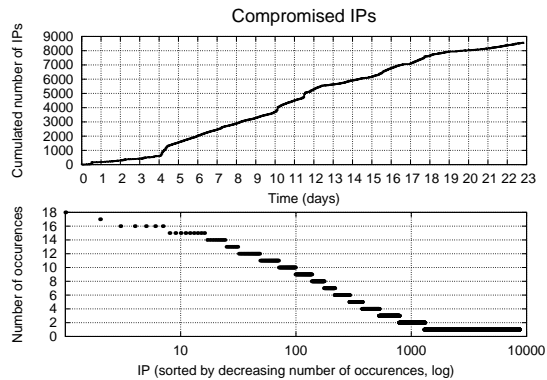}
  \caption{Number of unique IPs that we de-anonimized in time (top) and
    number of occurrences of a de-anonymized IP with at least one day of
    interval (bottom). \it{We compromised IPs uniformly in
      time. Whereas we de-anonymized most IPs once, a few IPs were
      de-anonymized many times.}}
  \label{fig:unique_IPs_compromised_in_time}
\end{figure}

We show the cumulative number of de-anonymized unique IP addresses in time in
Figure~\ref{fig:unique_IPs_compromised_in_time} (top). Apart from a few
bursts on days $4$, $10$, and $11$, we compromise IP addresses uniformly in time
with a rate of approximately $372$ IPs per day. As we instrument only
$6$ exit nodes, the cumulated number of de-anonymized IP addresses does not
converge yet after $23$ days.

However, we de-anonymize many IPs multiple times with at least one day
of interval in Figure~\ref{fig:unique_IPs_compromised_in_time}
(bottom). We have chosen at least one day of interval before
incrementing the number of occurrences of an IP to have a fair basis of
comparison among IPs. As the measurement lasts for $23$ days, the
maximum number of occurrences is $23$. We notice that the IP with the
largest number of occurrences appears $18$ times, meaning that we
de-anonymize that IP almost every day. This behavior suggests that
some IP addresses play a peculiar role in \bt{}, e.g., heavy downloaders, use Tor.

\subsection{Usage per Location}

We now correlate the location of the de-anonymized IP addresses of \bt{} users
on top of Tor to the location of $10$ million IPs of {\it regular} \bt{}
users. Those $10$ million IPs were collected on August $22$nd, $2009$
and are the merge of $12$ global snapshots of ThePirateBay, the
largest \bt{} tracker of the Internet, taken with an interval of two
hours. Although the practicality of collecting the IP of most \bt{}
users of the Internet is a serious privacy threat, the complete
description of this collect is beyond the scope of this paper. We
refer to~\cite{stevens_spying} for more information.

\begin{figure}[!ht]
  \centering
  \includegraphics[height=2in]{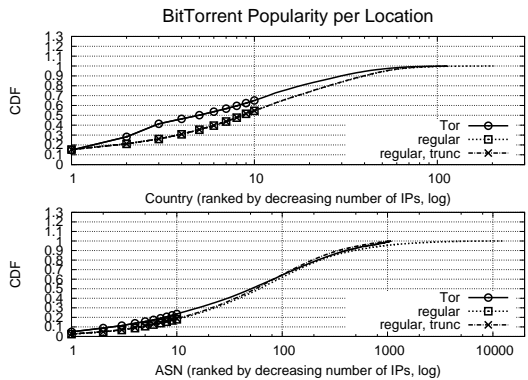}
  \caption{Distribution of IPs per country (top) and AS (bottom) for
    the \bt{} users on top of Tor (solid line) and the regular \bt{}
    users (hashed line). For both subfigures, we also show the CDF of users per country (resp. AS) in regular BitTorrent with the number of countries (resp. ASes) truncated to be the same as in Tor (dash-dot line).}
  \label{fig:IPs_per_Country}
\end{figure}

\textbf{Usage per Country.} Figure~\ref{fig:IPs_per_Country} depicts the distribution of the
de-anonymized IP addresses per country for both \bt{} users on top of Tor and
regular users. We observe that \bt{} users on top of Tor are
concentrated in fewer countries than regular \bt{} users. Indeed,
almost 70\% of them come from less than 10 countries, whereas in a
non-Tor environment, this same percentage is collected from around
$20$ countries. This is consistent with the fact that Tor is much more
popular in few countries as reported by \cite{Shining_Dark}. This is
illustrated in Table~\ref{tab:rank-top20-as} (left), that shows the
popularity of \bt{} on top of Tor per country. It is interesting to
note the ``Over'' column where we divide the fraction of users on top
of Tor in a given country by the fraction of regular \bt{} users in that
country. This clearly shows that comparatively to its regular usage,
\bt{} on top of Tor is extremely popular in UK, Japan, and
India. Besides Tor's popularity in different countries, the usage of
\bt{} on top of Tor might also be impacted by the severity of
copyright-laws in countries where \bt{} is used.

\begin{table} [!ht]
  \caption{Popularity of \bt{} on top of Tor per country (left) and AS
    (right). The over-representation (Over) for a given country
    (resp. AS) is the fraction of \bt{} IPs on top of Tor divided by
    the fraction of IPs on regular \bt{} for that country (resp. AS).}
  %\begin{minipage}{2in}
    \begin{center}
     % \tiny
      \begin{tabular}{|c|c|c|c|c|}
        \hline
        Rank & \# & \% & Over & CC\\
        \hline
        1  & 1,255 & 14 & 0.9 & US\\
        2  & 1,147 & 13 & 5.4 & JP\\
        3  & 1,125 & 13 & 2.8 & DE\\
        4  & 426   & 5  & 1.2 & FR\\
        5  & 321   & 3  & 1.3 & PL\\
        6  & 301   & 3  & 0.9 & IT\\
        7  & 264   & 3  & 0.7 & CA\\
        8  & 240   & 2  & 5.7 & IN\\
        9  & 232   & 2  & 0.9 & TW\\
        10 & 229   & 2  & 4.6 & UK\\
        \hline
      \end{tabular}
    \end{center}
  %\end{minipage}
 % \begin{minipage}{2in}

    \begin{center}
      %\tiny
      \begin{tabular}{|c|c|c|c|c|}
        \hline
        Rank & \# & Over & CC & AS\\
        \hline
        1  & 415 & 4.4 & DE & Deutsche Telekom\\
        2  & 338 & 5.5 & JP & NTT\\
        4  & 213 & 1.9 & MY & TMNet\\
        3  & 210 & 1   & IT & Telecom Italia\\
        5  & 156 & 0.9 & US & AT\&T\\
        6  & 148 & 1   & FR & Orange\\
        7  & 141 & 4   & DE & Hansenet\\
        8  & 134 & -   & CN & ChinaNet\\
        9  & 132 & 1   & PL & TPNet\\
        10 & 121 & 1.4 & FR & Free\\
        \hline
      \end{tabular}
  \label{tab:rank-top20-as}
    \end{center}

 % \end{minipage}
\end{table}

\textbf{Usage per AS.} Figure~\ref{fig:IPs_per_Country} (bottom) represents the distribution of IP addresses per
Autonomous System (AS) for both \bt{} users on top of Tor, and regular
users. We do not show the ``Over'' fraction for ChinaNet as we have
observed that Chinese users do not use ThePirateBay, the tracker that
we have used to capture the location of regular BitTorrent users.
Again, \bt{} users on top of Tor are concentrated in fewer ASes than
regular \bt{} users. This is consistent with the concentration into
few countries we have noticed previously. The table
\ref{tab:rank-top20-as} (right), representing the popularity of \bt{}
on top of Tor per AS, shows that comparatively to its regular usage,
\bt{} on top of Tor is extremely popular in NTT, HanseNet, and
Deutsche Telekom.

% From non-anon to non-priv
\section{From Non Anonymity to Privacy Issues}
\label{nonAnonymat_privacy}

Tor enables a user to anonymously browse the Internet, i.e., without revealing his IP address(es) to destinations. As such, Tor provides IP anonymity. It also prevents any entity from linking the source IP address and the destination IP address. It provides what we refer to as IP un-linkability.

In this section, we show that these properties are difficult to fulfill in practice. In particular, we show that \bt{} users take tremendous privacy risks while using Tor.
First, we present a coarse-grained analysis by focusing on de-anonymized users behavior during their web browsing sessions. Then, we show that an attacker can obtain, using the domino effect we described in Section~\ref{sec:attack_stats}, private data and link them to the de-anonymized user.

\subsection{Coarse-grained Analysis}

\begin{figure}[!t]
  \centering
  \includegraphics[height=2in]{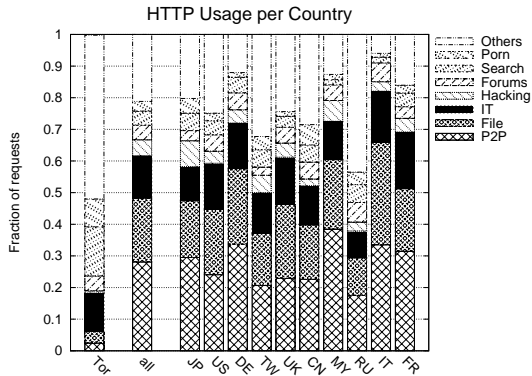}
  \caption{Histogram of the HTTP usage per country. {\it Tor}
    represents the overall distribution of requests per category for all Tor users. {\it all}
    represents the overall distribution of requests per category for de-anonymized users. The
    country codes represent the same information but for each of the
    top$10$ countries in number of requests.}
  \label{http_usage}
\end{figure}

Figure \ref{http_usage} is an illustration of the type of profiling that can be performed once the source IP addresses of Tor users are retrieved. It displays the ratio of most popular categories of sites accessed by Tor users according to their country of origin. These categories were selected using the classification made by \cite{classification}.

First, we notice that most of \bt{} users share a common ``typical'' behavior (independently from their origins): they are heavy downloaders. In fact, on average 50\% of their web usage is categorized as \texttt{Peer-to-Peer} or \texttt{File Sharing}, with slight variation noticed among different countries. This high frequency in accessing P2P web sites can be explained by the fact that since they are \bt{} users, they are often browsing torrents' search sites like \url{torrentz.com} or \url{mininova.com}. The \texttt{File Sharing} category is ranked as second, which shows that most \bt{} users \textit{also} use HTTP portals for file sharing and multimedia download. 

Second, as depicted in the bar labeled \texttt{Tor} in Figure \ref{http_usage}, typical\footnote{Referring to Tor users without any constraint on the protocols they are running.} Tor users' behaviors is significantly different from \bt{} users' typical web browsing. The latter have little interest in social networks or blog web sites (representing 13\% of the web sites common Tor users visited). On the other hand, \bt{} users seem more interested in forums and hacking sites. In the light of these observations, we can guardedly conclude that most of \bt{} users on top of Tor show higher IT skills than the average Tor users. Finally, it should be noted that Search, IT and Porn categories seem to be a constant in Tor typical usage. 

\subsection{Fine-grained Analysis: Users' Profiling}
In this section we show evidence of how anonymity issues may lead to privacy risks, while using \bt{} on top of Tor. Indeed, once de-anonymized, \bt{} streams can be linked to other applications' streams. Recall from Section~\ref{sec:attack_stats} that by exploiting the domino effect, we can link not only streams belonging to the same circuit to the de-anonymized IP address, but also other streams that may be associated with the same \bt{} user. This gives an adversary with valuable tools to (i) have a full list of torrent files that a targeted user is downloading, and more importantly to (ii) monitor, among others, the user browsing activities by sniffing the HTTP connections she is establishing through the controlled exit node.

As a consequence, the adversary can extract the user's visited web sites, retrieve searched keywords and even collect user's cookies that transit through the controlled exit node. If anonymized, this information may be not that important for any adversary. However, exploiting our attacks, and thus de-anonymizing the actual user's IP address originating the connections, such information become very sensitive, as several data mining techniques (e.g. \cite{data_mining1,data_mining2}) could be used to profile the targeted user.  

\begin{figure}[!t]
  \centering
  \includegraphics[height=2in]{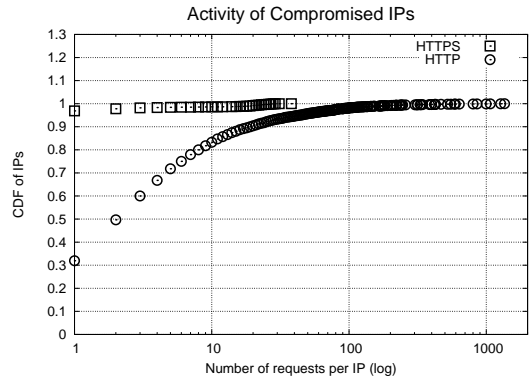}
  \caption{CDF of compromised IP addresses establishing HTTP/HTTPS connections.}
  \label{tor_http_req}
\end{figure}

An example of such serious risks \bt{} users are taking, when both using \bt{} and surfing the Internet, is illustrated by the high number of compromised IPs that were establishing HTTP connections in addition to \bt{}. Figure~\ref{tor_http_req} shows that roughly 70\% of users we compromised their IPs have used at least once both protocols on top of Tor at the same time. It also shows that several users are frequently browsing the Internet while using \bt{}, with more than 100 established HTTP connections we linked to the compromised IP.

The risks these users are taking are even more important when the Internet activity is originated from countries known to have ``restrictive'' Internet access. As an illustrative example, and for the purpose of proving the sensitive information that can be collected using our de-anonymizing techniques, we have identified and profiled several IPs from China and Myanmar frequently accessing web sites and blogs that belong to political opposition groups. This evidence shows how dangerous our attacks could be, especially if used by third parties to profile users and track ``deviant'' Internet access, infringing then private users' data.

% Conclusion
\section{Discussion and Conclusions}

We have presented three techniques targeting the anonymity of \bt{} users on top of Tor. In practice, we have demonstrated how an adversary may, with low resources, break users anonymity and shown evidence of serious privacy risks this might induce. We also described, through a so-called domino effect, how identity leak may contaminate different protocols on top of Tor. 

In addition, we have quantified and characterized the \bt{} usage on top of Tor. Exploiting our de-anonymizing attacks, this paper shows the disconnection between users' willingness to use \bt{} anonymously and their expectation to preserve their identity through Tor. In essence, even if \bt{} users expect Tor to provide  anonymity and IP un-linkability, we show that this is not actually the case. In other words, \bt{} users are in general not more protected on top of Tor than elsewhere. Our findings may then discourage \bt{} users from using the Tor network, freeing it from an useless (and undesirable), yet important load.

Even though a solution consisting in end-to-end encryption and authentication in \bt{} might countermeasure our attacks, we believe this would be a costly solution for trackers to implement, and would induce higher latencies into \bt{} connections. These non desirable properties such solution exhibits would certainly make heavy downloaders and content providers reluctant to adopt it.

In summary, two factors help our attacks to succeed. First, because \bt{} is used on top of Tor, it becomes more vulnerable to traffic monitoring and even to communications' hijacking. Second, the lack of cryptographic material's support among \bt{} entities creates a gap between security and users' expectations when using Tor. We believe this security versus performance balance (also well illustrated by Google persisting in not using HTTPS for a vast majority of its services despite serious risks~\cite{letter}) should be carefully considered not only by \bt{} content's providers, but also by users that are willing to anonymously use both \bt{} and other protocols on top of Tor.

%
% ---- Bibliography ----
%

%\small{
\bibliographystyle{plain}
\bibliography{biblio}
%}
%
\end{document}